\documentclass{article}

\usepackage{PRIMEarxiv}

\usepackage[utf8]{inputenc} 
\usepackage[T1]{fontenc}    
\usepackage{hyperref}       
\usepackage{url}            
\usepackage{booktabs}       
\usepackage{amsfonts}       
\usepackage{nicefrac}       
\usepackage{microtype}      
\usepackage{lipsum}
\usepackage{fancyhdr}       
\usepackage{graphicx}       
\graphicspath{{media/}}     
\usepackage{amsmath}

\title{KDPrint: Passive Authentication using Keystroke Dynamics-to-Image Encoding via Standardization
}

\author{
  Yooshin Kim, Donghoon Shin \\
  dept. Electrical Eng. and Computer Sci. \\
  DGIST \\
  Daegu, Korea\\
  \texttt{\{yooshin0303, dshin\}email@email} \\
    \And
  Namhyeok Kwon \\
  School of Undergraduate Studies \\
  DGIST \\
  Daegu, Korea\\
  \texttt{knh7345@dgist.ac.kr} \\
}

\begin{document}
\maketitle

\begin{abstract}
In contemporary mobile user authentication systems, verifying user legitimacy has become paramount due to the widespread use of smartphones. Although fingerprint and facial recognition are widely used for mobile authentication, PIN-based authentication is still employed as a fallback option if biometric authentication fails after multiple attempts. Consequently, the system remains susceptible to attacks targeting the PIN when biometric methods are unsuccessful. In response to these concerns, two-factor authentication has been proposed, albeit with the caveat of increased user effort. To address these challenges, this paper proposes a passive authentication system that utilizes keystroke data, a byproduct of primary authentication methods, for background user authentication. Additionally, we introduce a novel image encoding technique to capture the temporal dynamics of keystroke data, overcoming the performance limitations of deep learning models. Furthermore, we present a methodology for selecting suitable behavioral biometric features for image representation. The resulting images, depicting the user's PIN input patterns, enhance the model's ability to uniquely identify users through the secondary channel with high accuracy. Experimental results demonstrate that the proposed imaging approach surpasses existing methods in terms of information capacity. In self-collected dataset experiments, incorporating features from prior research, our method achieved an Equal Error Rate (EER) of 6.7\%, outperforming the existing method's 47.7\%. Moreover, our imaging technique attained a True Acceptance Rate (TAR) of 94.4\% and a False Acceptance Rate (FAR) of 8\% for 17 users.
\end{abstract}


\section{Introduction}
With the increasing usage of smartphones\cite{b1}, it has become crucial for users to determine the legality of their actions in the mobile environment. Modern mobile user authentication systems are broadly categorized into screen-based and biometric authentication. Screen-based authentication typically includes personal identification numbers (PINs) and patterns, while biometric authentication encompasses fingerprint and face recognition. Issues stemming from the vulnerabilities of PIN authentication\cite{b2}, such as smudge attacks and shoulder surfing, have prompted the emergence of authentication systems based on user-specific properties like fingerprints or faces. However, these physical biometric authentication methods are unstable due to their immutable nature. These characteristics could be influenced by surrounding environments and altered by temporary scars, leading to incorrect results in the system. Additionally, biometric systems suffer from the drawback of irreusable unique information in case of leakage. Furthermore, the system is still vulnerable to attacks targeting PINs, as a backup PIN authentication system is provided if the maximum authentication attempts for biometric system fail. Therefore, in the event of a mobile device being stolen, there is a possibility of personal information leakage through the final security system, PIN authentication.

To address such issues, two-factor authentication has been proposed. Two-factor authentication is an identity and access management security method that requires two forms of authentication when accessing resources and data. This typically involves utilizing hardware tokens, short message service (SMS) authentication, and more. For instance, when authenticating identity on a computer, the user initially inputs a password, and subsequently, enters a token sent to their smartphone via SMS or email. While two-factor authentication provides higher security than one-factor authentication, it compromises user usability by requiring additional steps\cite{b30}, which can discourage user adoption. 

To overcome this limitation, research has investigated passive authentication, a technique for verifying users without necessitating additional actions from them, alongside keystroke dynamics, a biometric-based user authentication method grounded in human behavior. Keystroke dynamics assesses user typing patterns and behaviors, recognizing distinctive keystroke patterns based on key sequence, intervals, press intensity, and press duration\cite{b3}, while passive authentication validates user identity without user involvement. Utilizing keystroke dynamics-based authentication systems can enhance security without compromising user usability. However, it is premature to replace traditional authentication systems due to technical issues. Hence, this research endeavors to leverage this technology to enhance security in scenarios demanding high-level security, through the implementation of keystroke dynamics as a passive authentication method.

The recent advancements in deep learning suggest the potential to learn unique patterns in user behavior in keystroke dynamics\cite{b7, b4,b5,b6,b8,b10,b11}. Currently, deep learning provides state-of-the-art results in a supervised setup, where the model is trained on data that sufficiently represents both genuine users and imposters. However, defining and collecting all possible types of genuine and imposter data is impossible. Therefore, there is a need to devise methods for anomaly detection using only information collected under normal circumstances. In the case of PIN authentication, frequently used PIN tend to be fixed over the duration of smartphone usage, forming patterns unconsciously in their input. Previous research has demonstrated the effectiveness of utilizing deep learning models to address this issue\cite{b4,b5,b6,b8,b10,b11}. However, recent progress in deep learning has primarily concentrated on image processing\cite{b12,b13}. Specifically, a novel method leveraging the latest advancements in deep learning has been suggested to convert time-series data into images\cite{b14, b24, b25, b26, b27, b28}. This transformation is beneficial for highlighting or compacting temporal information and hidden patterns within the data, aiding the model in detecting anomalies.

In this paper, we propose a novel framework for performing passive authentication using keystroke dynamics encoded into images, utilizing touch location data not commonly employed in traditional datasets. Additionally, assuming an unsupervised setup, we compare their high-performance method of transforming time-series data into images with the proposed approach. Our main contributions are as follows:

\begin{itemize}
    \item We propose a novel approach using keystroke dynamics-to-image encoding for malicious user detection. This also serves as a method for passive authentication, enhancing user convenience and expanding applicability.
    \item We collect a keystroke dynamics dataset and pre-process it using standard normalization instead of min-max normalization. This results in greater versatility of the impostor image compared to the genuine user image.
    \item The proposed keystroke dynamics-to-image encoding excels in anomaly detection, outperforming baselines on our dataset with simple one-class classification model. This highlights the superior efficiency of our encoding method in conveying keystroke dynamics information.
\end{itemize}

The remainder of the paper is organized as follows. Section \uppercase\expandafter{\romannumeral2} explores related studies, Section \uppercase\expandafter{\romannumeral3} outlines our framework, detailing the collected data and providing specific descriptions of our efforts and Section \uppercase\expandafter{\romannumeral4} reports experimental results. Sections \uppercase\expandafter{\romannumeral5} and \uppercase\expandafter{\romannumeral6} address discussion and conclusion, respectively.

\section{Related Work}

In this section, we outline several key steps for biometric authentication and explore research efforts focused on transforming time-series data into images for enhanced analysis and recognition.

Implementing a keystroke dynamics system involves key steps for effective authentication\cite{b3}. It starts with \textbf{\textit{data collection}}, recording users' keystroke patterns, crucial for subsequent analysis. \textbf{\textit{Feature extraction}} follows, identifying and extracting relevant features from the keystroke data. \textbf{\textit{Classification}} algorithms then analyze these features to differentiate between legitimate users and imposters, ensuring robust and reliable authentication.

\textbf{\textit{Dataset}} The conventional approach has been to collect a timing value-based feature set to utilize both computers and mobile devices \cite{b3}. However, specific features tailored for mobile devices, such as touch functionality, can enhance user profiling \cite{b15}. Selecting an appropriate feature set is challenging as datasets are either not disclosed due to privacy concerns or vary across studies. Although publicly available datasets like CMU\cite{b16}, Clarkson\cite{b17}, Buffalo\cite{b18}, and HMOG\cite{b19} exist, they do not include touch features. Hence, this study gathered data to utilize touch features for analysis.

\textbf{\textit{Feature Extraction}} Upon completion of data collection, the subsequent step involves feature extraction to identify users using classification algorithms. A prominent feature extraction method based on timing values includes five distinct categories, divided as monographs and digraphs. Monographs consist of hold duration time, while digraphs involve key down to down time (DD), key up to down time (UD), key up to up time (UU), and key down to up time (DU), all derived from the key pressed time and released time of keystrokes\cite{b3}.

\textbf{\textit{Classification}} Based on such data, recent advancements in machine learning (ML) and deep learning (DL) technologies have led to improved performance in authentication systems. In \cite{b10}, a randomly generated keypad and support vector machine (SVM) were employed to achieve an equal error rate (EER) of 9.67\% on acquired keystroke data. Meanwhile, \cite{b20} utilized SVM to achieve an accuracy of 97\%, focusing on learning differences among users. Efforts to apply DL have evolved from multilayer perceptron (MLP) networks \cite{b4, b8} to recurrent neural networks (RNN) and long short-term memory (LSTM) networks, aiming to leverage the temporal characteristics of keystroke dynamics. In \cite{b11}, LSTM was used on keystroke and sensor data from mobile devices, achieving an EER of 4.62\%. While these approaches better represent users than ML, they are model-dependent and lack data processing, leaving room for performance improvement.

\textbf{\textit{Time-series to Image}} Keystroke dynamics are primarily represented using time-series data. Recently, there has been significant interest in converting such data into images\cite{b14, b21}. This transformation is commonly utilized in DL models specialized in image classification to enhance model performance. The conversion from time-series to images aims to visualize hidden pattern, facilitating more effective recognition and analysis by the model. This approach opens up new possibilities in keystroke dynamics. In pioneering work by\cite{b22}, recurrence plots (RP) were introduced to capture recurrence patterns in time-series data. Additionally, encoding methods like Gramian angular fields (GAF) and Markov transition fields (MTF) have been proposed for representing time-series data\cite{b23}. Recently, there has been a trend towards representing the extracted features in an image-like matrix using keystroke dynamics data\cite{b24,b25,b26,b27,b28}. Despite achieving high performance, a limitation persists in that it heavily relies on the model's performance without substantial emphasis on imaging methodology. Accordingly, our research prioritizes the visualization of keystroke dynamics data through the utilization of simple models, rather than emphasizing model performance.

\section{System Methodology}

This section provides an overview of the overall system, including a detailed description of the collected dataset, the process of converting keystroke dynamics into images, and the unsupervised learning approach employed for authenticating mobile device identities. In this research, our objectives are twofold: 1) to gather the data without additional equipment to ensure expandability, and 2) to utilize this data to generate images that exhibit discernible distinctions between keystroke images of genuine user and imposters, thereby facilitating passive authentication of user login activity. The model is trained in an unsupervised manner using image representations of time series features extracted from keystroke dynamics data. This approach enables the model to detect imposters by representing the genuine user's feature space.

\begin{figure}[htbp]
\centerline{\includegraphics[width=0.5\textwidth]{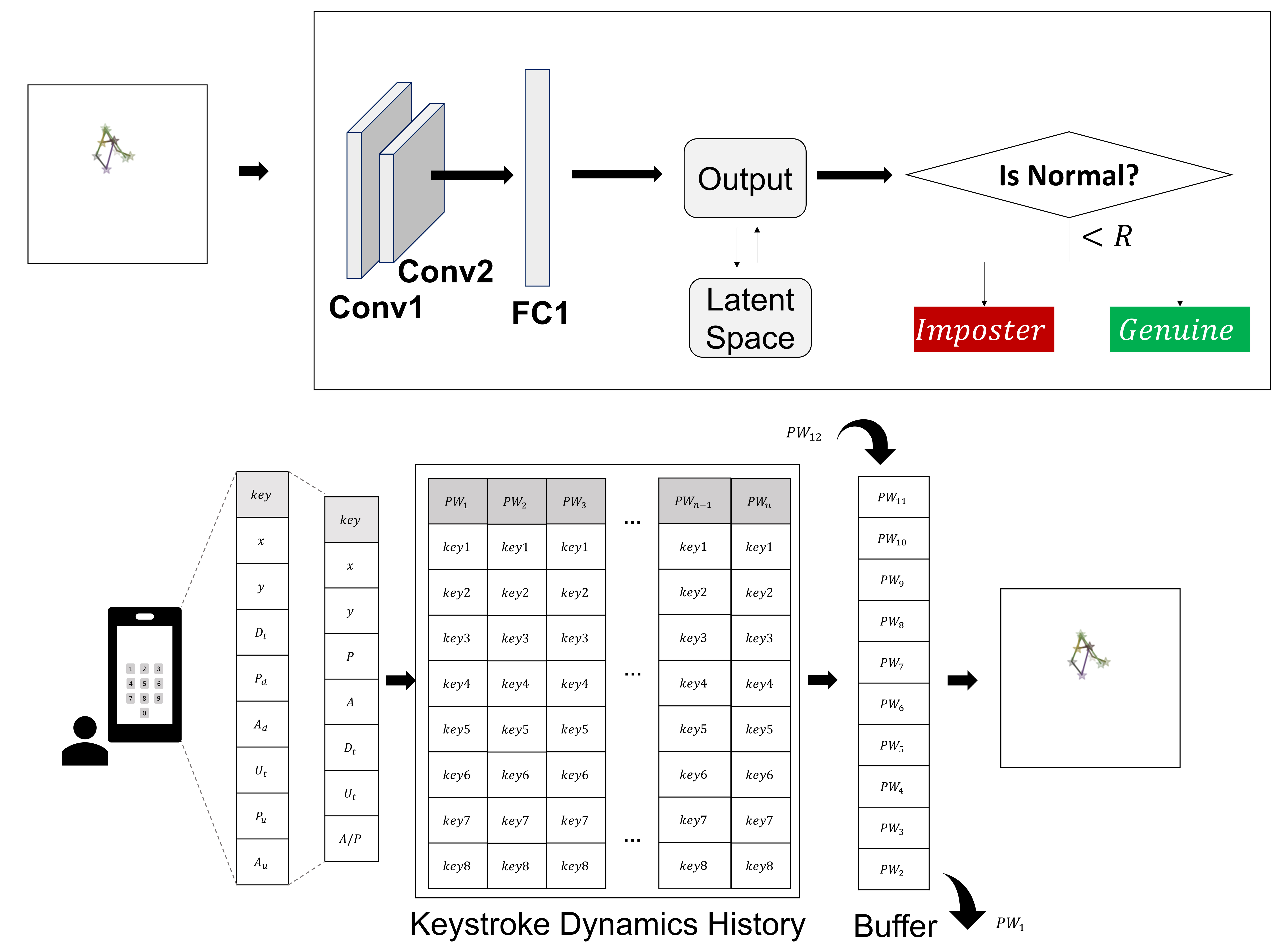}}
\caption{The overall user authentication process of the proposed system.}
\label{fig:Overall}
\end{figure}

\subsection{System overview}

Our system serves as a secondary authentication method utilized for user authentication in activities requiring strong security on mobile devices. When keystroke dynamics data from PIN authentication is collected, an image representing the user's input pattern is generated. The criteria for image generation lie in the noticeable differences between genuine user and imposters. Due to the instability of keystroke patterns, we use a weighted buffer to remove noise from each input and generate images with sensitivity to recent inputs. The generated images are then trained using a simple one-class classification model, deep support vector data description (SVDD) \cite{b29}. A simple model is used to demonstrate the effectiveness of the proposed image encoding method. During the training phase, the network learns the user's representation based on randomly generated latent vectors. Our system aims to detect attackers in PIN authentication scenarios where the attacker knows the user's PIN and attempts to steal the user's mobile device to gather high-security level information. If the system detects suspicious authentication attempts, it alerts the device administrator or requests strong authentication methods such as two-factor authentication. The overall process of the proposed system is illustrated in Fig. \ref{fig:Overall}.

\subsection{Dataset}

\begin{figure}[htbp]
\centerline{\includegraphics[width=0.49\textwidth]{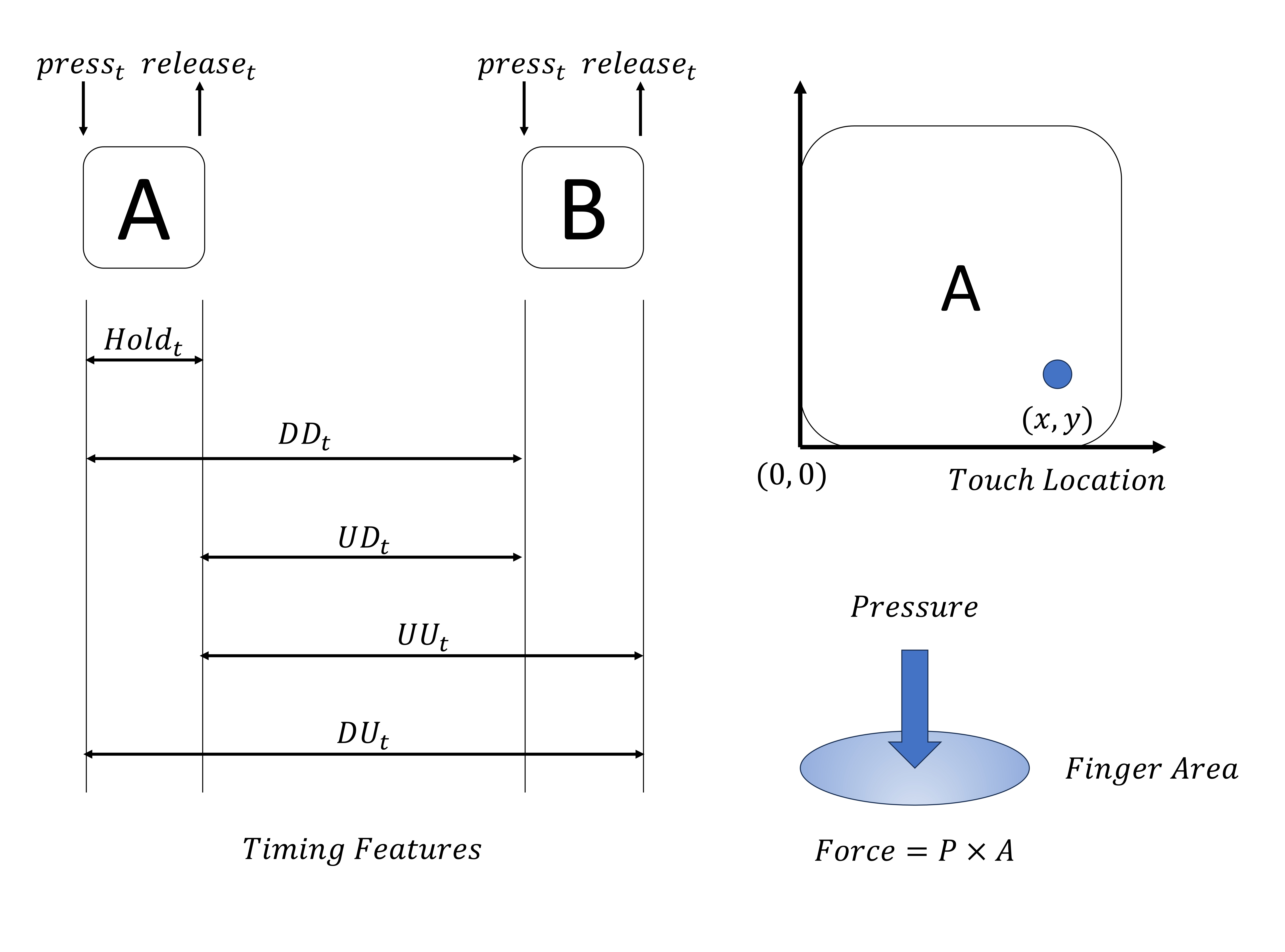}}
\caption{The feature set used in out dataset.}
\label{fig:feature}
\end{figure}

Constructing appropriate features in a keystroke dynamics-based authentication system is crucial as it determines the system's robustness and the users' expressiveness. Timing data, primarily collected in traditional datasets, are susceptible to tampering, leading to instability in authentication\cite{b31}. On the other hand, touch location data are difficult to acquire or tamper with unless hidden applications are installed, and they have shown potential in enriching users' expressiveness\cite{b15}. Therefore, our data collection application records press and release times, pressure, finger area, and touch location, all of which are obtainable using Android Studio without the need for additional equipment. From this data, timing features, touch location features, and force features are derived. The extractable features in our dataset are depicted in Fig. \ref{fig:feature}. Timing features include hold duration time, key down to down time (DD), key up to down time (UD), key up to up time (UU), and key down to up time (DU). Touch location indicates the fingertip's position within the keypad buttons. Lastly, force features are calculated as the product of pressure and area. In scenarios requiring high security, these keystroke features are logged and analyzed to ascertain the user's identity upon entering a PIN.

\subsection{Feature engineering}

The collected data spans multiple domains, which can hinder the learning process of deep learning algorithms. Therefore, data pre-processing is conducted to adjust the scale and identify the user's hidden keystroke patterns. One commonly used method is min-max normalization, which scales the data to a range of 0 to 1 while highlighting outliers. However, in keystroke dynamics data where each input contains noise, min-max normalization is not suitable as the noise can be treated as outliers. Additionally, upon analyzing the collected dataset, it was found that each data follows a normal distribution, indicating that standard normalization is more appropriate. Therefore, the collected data is scaled using standardization to remove noise and stored in a buffer.

\begin{equation}
    \text{Buffer}_{t} = \frac{1}{2(B-1)} [\sum_{{i=t-B+1}}^{t-1} kd_i + (B-1)kd_t]\label{eq:quadratic}
\end{equation}

The pre-processed data is stored in the buffer according to the input order, as described in \eqref{eq:quadratic}. The buffer holds up to \(B\) user inputs, and their averages are calculated to reduce data noise. This approach aims to learn more stable and accurate hidden patterns of the user by offsetting individual noises through the averaging of multiple inputs. However, imposter inputs may not be correctly detected due to the influence of normal user inputs stored in the buffer. For this reason, a weighted mean is used in the buffer. By assigning a weight of \(B-1\) to the most recent input, it adjusts the balance with the data stored in the buffer, thereby effectively detecting impostor inputs while managing normal user noise.

Considering that the collected dataset may not sufficiently represent the user's input pattern, the inadequate information in the training dataset raises concerns about the model's performance in accurately identifying genuine users. To overcome this issue, the quantity of data is increased using the buffer. The total number of unique combinations that can be generated by the buffer, given \(N\) instances per user, becomes \( \binom{N}{B} \) possible instances. These data augmentation in training enhances the representational power of the dataset, thereby improving model performance.

\subsection{Image generation}

\begin{figure}[htbp]
\centerline{\includegraphics[width=0.5\textwidth]{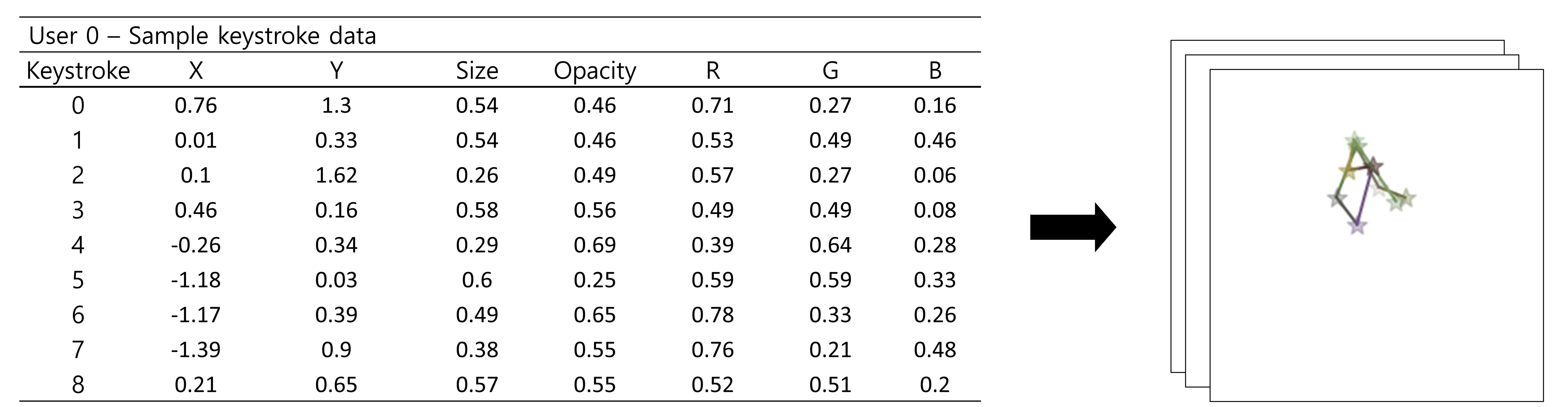}}
\caption{The example of feature-engineered dataset table and image generated using our proposed method.}
\label{fig:encoding}
\end{figure}

\begin{figure}[htbp]
\centerline{\includegraphics[width=0.38\textwidth]{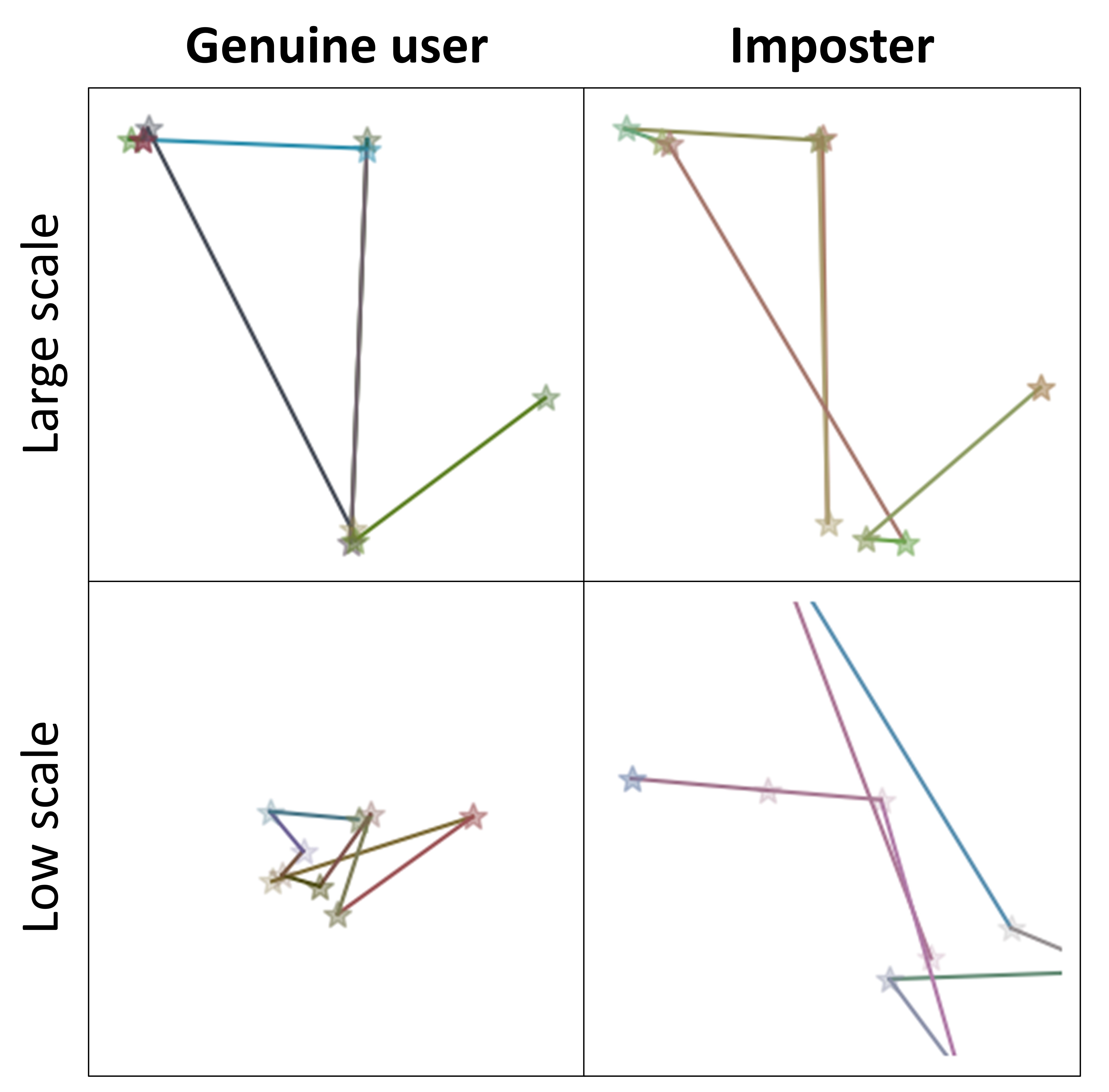}}
\caption{The keystroke dynamics to image encoding result with different scale.}
\label{fig:scale_comparison}
\end{figure}

The keystroke dynamics data, after completing the pre-processing stage, undergoes a process of conversion into images as shown in Fig. \ref{fig:encoding}. The goal of transforming keystroke data into images is to represent temporal information that cannot be observed in the original data. In a naive approach, circles of finger area size can be drawn at the user's touch coordinates on the keypad's display area. However, since our attacker knows the user's PIN, representing touches on a whole keyboard screen makes it difficult to discern features. Therefore, we upscale the screen as shown in Fig. \ref{fig:scale_comparison}, representing the touches on a single keypad button instead of the entire keyboard display area. By directly depicting the user's touch on the keypad button, more detailed touch analysis becomes possible compared to representing the overall touches.

\begin{figure}[htbp]
\centerline{\includegraphics[width=0.5\textwidth]{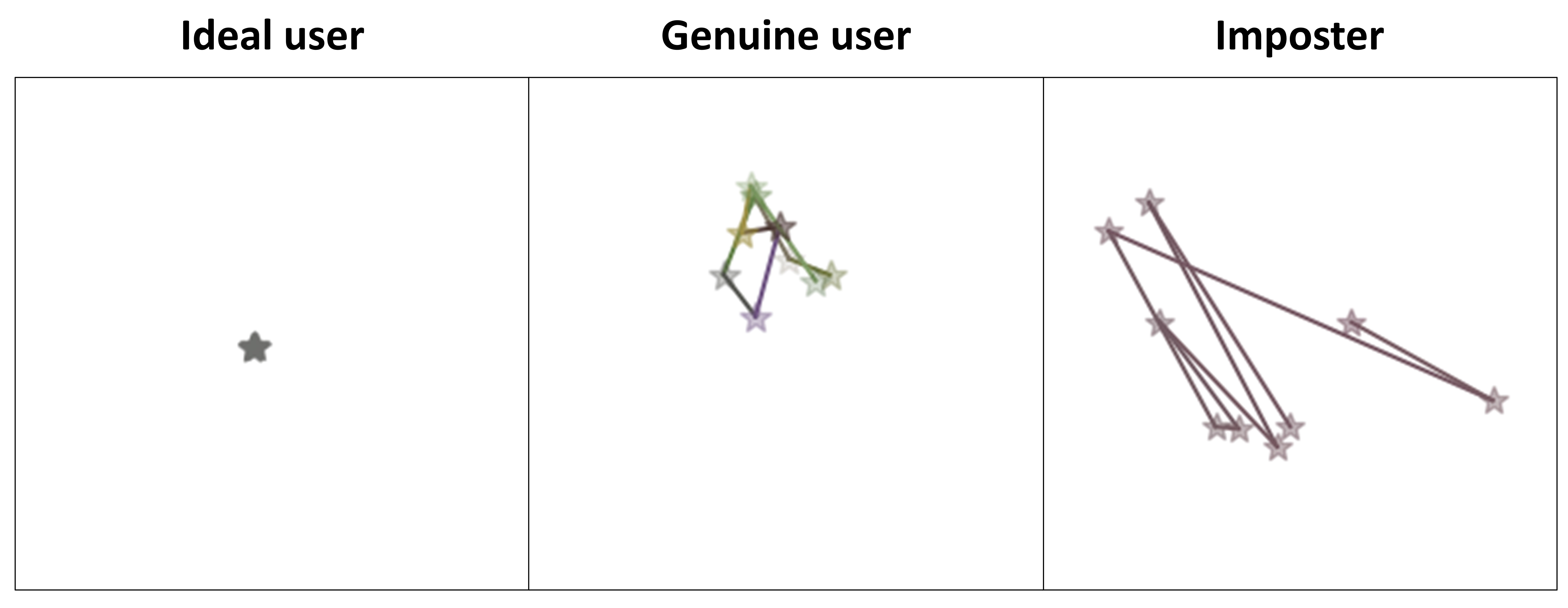}}
\caption{The results of image encoding for the ideal, real, and impostor cases.}
\label{fig:idealuserinput}
\end{figure}

The proposed method for image generation involves representing all keystroke information on a single keypad button. The essential data types required for generating these images encompass Position, Marker Size, Color, and Opacity. Position signifies the touch location within the key and is represented using standardized (x, y) data. Marker size indicates the size of the marker (*) at that position. Color represents the color of the marker, with three channels for red, green, and blue. Lastly, Opacity indicates the transparency of the marker. Due to the standardization of data, in the ideal user input, all keystrokes converge to the origin, resulting in a stabilizing effect. Consequently, the image of a legitimate user tends to exhibit greater stability compared to that of an impostor, as illustrated in Fig. \ref{fig:idealuserinput}.

The configuration elements of the five features of encoding, excluding Position, are determined by applying principal component analysis (PCA) to the dataset excluding (x, y). The transformation matrix is determined using the eigenvalues of the dataset when the coverage ratio for each keystroke exceeds 90\%. Min-max normalization is applied to restrict the range of individual components for every keystroke. Since the image is primarily influenced by the Position, the eigenvalues of maximum magnitude are utilized sequentially to determine the radius, opacity, and color of the marker, thereby enhancing image stability.

\subsection{Deep SVDD}

One-class classification has undergone significant research, from the introduction of the one-class support vector machine (OC-SVM) to the evolution of support vector data description (SVDD). In order to enhance the representation of data features more effectively than traditional methods, the concept of deep SVDD, built upon the architecture of deep neural networks, has been proposed \cite{b29}. Similar to the kernel in SVDD, deep SVDD aims to utilize deep learning to identify a more distinct hyperspace within the feature space.

The objective of SVDD is to find the smallest sphere in the feature space that contains normal data with a center \(c\) and a radius \( R \). The goal is to locate the sphere's boundary such that it encompasses the normal data points. The objective function is defined as follows.

\begin{equation}
\begin{aligned}
\min_{R, c, \epsilon} \quad R^2 +& \frac{1}{v_n} \sum_{i} \epsilon_i \\
\text{s.t.} \quad \|k(x_i) - c\|_F^2 \leq R^2 &+ \epsilon_i, \quad \epsilon_i \geq 0 \quad \text{for all } i
\end{aligned}
\end{equation}

Utilizing the slack variable, the soft boundary is adjusted, and \(v\) is employed to regulate the trade-off between the slack variable and \(R\).

In deep SVDD, the goal is to find the smallest sphere with a center point \(c\) and radius \(R\) by learning a deep learning weight \(W\) that represents data in a feature space. The objective function of simplified deep SVDD for learning is as follows.

\begin{equation}
\min_{W} \frac{1}{n} \sum_{i=1}^{n} \left\| (x_i;W) - c \right\|_2^2 + \sum_{l=1}^{L} \left\| W_l \right\|_F^2
\end{equation}

The anomaly score \(s(x)\) for detection is defined as follows. As can be seen from the objective function, the distance from the center of the sphere is used.
\begin{equation}
s(x) = \lVert (x; W^*) - c \rVert_2 
\end{equation}

The anomaly is detected through the difference between anomaly score \(s(x)\) and the radius \(R\) of the sphere. The decision value is defined by \(f(x)=s(x)-R\). In prediction, given a test data with generated image \(x_t\), if \(f(x_t)\) is larger than zero, we predict it as positive, otherwise we predict it as negative.

\section{Experimental Results}

In this section, we provide an explanation of the evaluation criteria used for assessment and discuss the results for the collected dataset. We also elaborate on the performance lower boundary obtained through the baseline ML and DL algorithms for anomaly detection. Additionally, we present experimental results comparing the performance of our system with existing researches.

\subsection{Performance metrics}
In this paper, we will describe the model's performance and experimental results using the following five performance metrics. The utilized performance metrics include false acceptance rate (FAR), false rejection rate (FRR), true acceptance rate (TAR), accuracy (ACC), and equal error rate (EER). FAR represents the ratio of imposter challenges incorrectly accepted as genuine user. FRR indicates the rate at which the model incorrectly identifies actual user as imposters. TAR, the inverse of FRR, represents the rate at which the actual user is accurately recognized and allowed. Accuracy denotes the proportion of correctly classified samples out of the total authentication tests, reflecting the overall performance of the model. EER is crucial for evaluating the performance of binary classification models, signifying the point where FAR and FRR become equal; a lower EER value indicates higher model performance.

\begin{itemize}
    \item \(FAR = FA / (FA + TR)\)
    \item \(FRR = FR / (FR + TA)\)
    \item \(TAR = TA / (TA + FR)\)
    \item \(Accuracy = (TA + TR) / (TA + TR + FA + FR)\)
    \item \(EER\): The point at which FAR and FRR are equal.
\end{itemize}

\subsection{Dataset}

\begin{figure}[htbp]
\centerline{\includegraphics[width=0.3\textwidth]{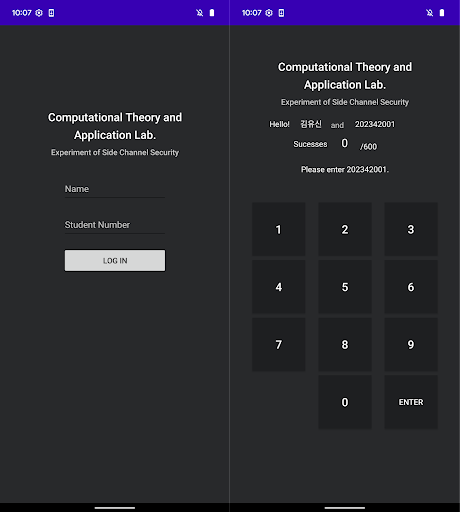}}
\caption{The application screens for collecting the dataset.}
\label{fig:app}
\end{figure}

An Android application running on the Google Pixel 6a device was utilized to collect keystroke dynamics data. The gathered data is stored locally in separate files to ensure that errors in one file do not affect others. The application includes a login screen, similar to Figure \ref{fig:app}, where users must input their name and Student ID for authentication. The Student ID is a unique 9-digit number specific to each student and does not overlap. Upon successful login, users are presented with a numeric keypad to input their PIN. Once they enter their Student ID and press the ENTER button, the data is saved locally. To prevent inaccurately entered data, the application does not collect data that deviate from the logged Student ID. Since the application was developed using Android Studio, no additional equipment is required to utilize our system.

We recruited 17 participants with an average age of 21 years (range: 20-24). Among these participants, 14 were male and 3 were female, all of whom were either undergraduate or graduate students. They reported predominantly using smartphones and typing with their right hand. The experiment was conducted in a laboratory setting, where participants were seated comfortably in chairs.

The experiment consisted of 6 sessions, each comprising 100 PIN entries. Participants paused their input at the 1st, 10th, and 100th entries for 1, 10, and 30 seconds, respectively, to prevent habituation to the repetitive task. The imposter's data, collected within subjects, involved entering the Student ID of different participants once each. The experiment lasted approximately 60 minutes per participant, and they received compensation of 10 dollars for their participation.

Since the users utilized a device they have never used before, we exclude data from the first session. From the data collected through our application, we generated data for 17 individuals, comprising 10,200 normal instances, and 170 imposter instances. Through data augmentation, we increased the number of instances, creating 2,400 training instances for each user, along with validation and test datasets consisting of 800 normal and imposter instances for each. The possible permutations of augmentation are determined by \( \binom{N}{B} \) and the total number of dataset, \(N\) was experimentally set to 4,000. To assess the proper formation of the collected dataset, we performed anomaly detection using both ML and DL on our raw dataset. These results also serve as ground truth to evaluate the performance of our keystroke dynamics to image conversion.

\begin{table}[htbp]
\caption{ Anomaly detection results of ML and DL algorithm on our dataset pre-processed by standardization.}
\begin{center}
\begin{tabular}{|l|l|l|l|l|l|}
\hline
Model & EER & FAR & FRR & TAR & ACC \\
\hline
Isolation Forest & 0.084 & 0.086 & 0.081 & 0.918 & 0.915 \\
OC-SVM & 0.083 & 0.082 & 0.084 & 0.915 & 0.916 \\
Autoencoder & \textbf{0.072} & \textbf{0.073} & \textbf{0.071} & \textbf{0.928} & \textbf{0.927} \\
\hline
\end{tabular}
\label{tab:rawdatasetresult}
\end{center}
\end{table}

The evaluation algorithm includes isolation forest (IF), OC-SVM, and autoencoder (AE). The detection threshold was determined at the EER on the validation dataset. For the autoencoder, the number of epochs is set to 100, and the latent space dimension is 30. The accuracy and other metrics are summarized in Table \ref{tab:rawdatasetresult}. Our objective is to achieve performance that is at least similar to or better than the performance of these algorithms.

\subsection{Our system performance}

Deep SVDD is applied with the following configurations for imposter detection in our system. The network consists of three layers, including two convolution layers and one fully connected layer. The learning rate is set to 0.001, the number of epochs is 200, and the dimension of the latent space is 64. Adam was used as the optimizer and the detection threshold was determined at the EER on the validation dataset.

\begin{table}[htbp]
\caption{The imposter detection results of autoencoder and ours.}
\begin{center}
\begin{tabular}{|l|l|l|l|l|l|}
\hline
Model & EER & FAR & FRR & TAR & ACC \\
\hline
Autoencoder &  0.072 &  0.073 &  0.071 &  0.928 &  0.927 \\
OURS(x,y) & 0.097 & 0.113 & 0.080 & 0.919 & 0.902 \\
OURS(PCA) & \textbf{0.067} & \textbf{0.080} & \textbf{0.055} & \textbf{0.944} & \textbf{0.932} \\
\hline
\end{tabular}
\label{tab:ourresult}
\end{center}
\end{table}

\begin{table}[htbp]
\caption{ OURS(PCA) authentication result for each user.}
\begin{center}
\begin{tabular}{|l|l|l|l|l|l|}
\hline
User & EER & FAR & FRR & TAR & ACC \\
\hline
user0& 0.029& 0.051& 0.006& 0.994& 0.971\\
user1& 0.069& 0.070& 0.069& 0.931& 0.931\\
user2& 0.085& 0.056& 0.114& 0.886& 0.915\\
user3& 0.119& 0.140& 0.098& 0.902& 0.881\\
user4& 0.050& 0.040& 0.060& 0.940& 0.950\\
user5& 0.041& 0.070& 0.011& 0.989& 0.959\\
user6& 0.052& 0.068& 0.036& 0.964& 0.948\\
user7& 0.013& 0.010& 0.016& 0.984& 0.987\\
user8& 0.057& 0.083& 0.031& 0.969& 0.943\\
user9& 0.114& 0.225& 0.003& 0.998& 0.886\\
user10& 0.125& 0.117& 0.133& 0.867& 0.875\\
user11& 0.155& 0.156& 0.154& 0.846& 0.845\\
user12& 0.049& 0.039& 0.059& 0.941& 0.951\\
user13& 0.022& 0.022& 0.022& 0.978& 0.978\\
user14& 0.089& 0.136& 0.041& 0.959& 0.911\\
user15& 0.034& 0.021& 0.046& 0.954& 0.966\\
user16& 0.052& 0.061& 0.043& 0.958& 0.948\\
\hline
Average& 0.067 & 0.080 & 0.055 & 0.944 & 0.932 \\
Min & 0.013 & 0.010 & 0.003 & 0.846 & 0.845 \\
Max & 0.155 & 0.225 & 0.154 & 0.998 & 0.987 \\
\hline
\end{tabular}
\label{tab:ourallresult}
\end{center}
\end{table}

Our system achieved an accuracy of 93.2\%, an EER of 0.06, and a FAR of 0.08 for 17 subjects on average, while the accuracy of the baseline with the raw dataset was 92.7\%. These results signify that transforming pre-processed keystroke dynamic data into images through standardization is effective for impostor detection. Additionally, we present results when using only touch location data and when combining it with PCA-transformed data, demonstrating the appropriateness of the PCA selection method. By employing PCA, we observed a performance improvement of 3\% in both EER and accuracy. The imposter detection results are presented in Table \ref{tab:ourresult}, and the detailed results for each user are described in Table \ref{tab:ourallresult}.

\subsection{Comparison with other time-series to image encoding}

\begin{figure}[htbp]
\centerline{\includegraphics[width=0.43\textwidth]{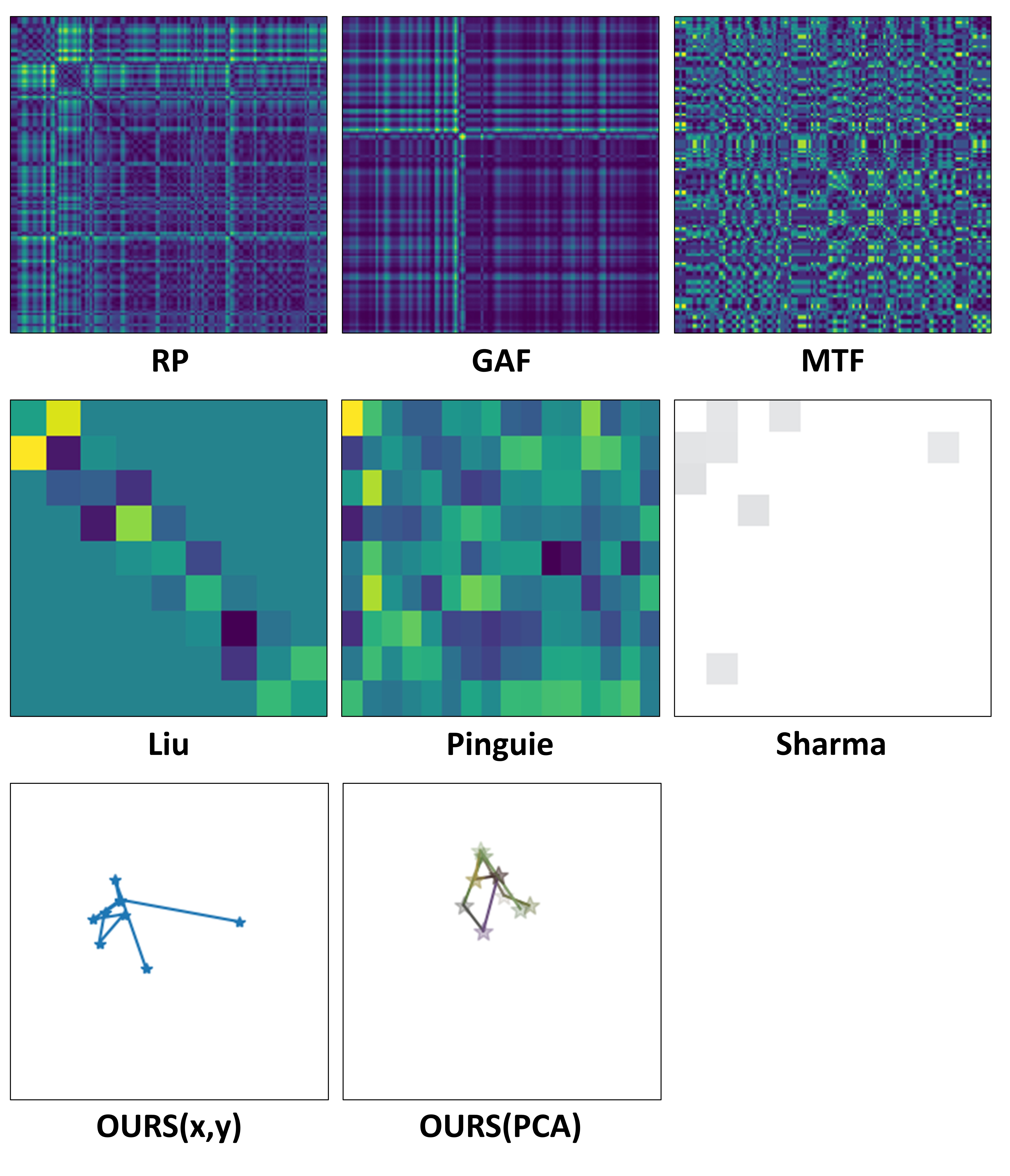}}
\caption{The implementation result of previous researches and ours.}
\label{fig:allimages}
\end{figure}

The image encoding methods used as baselines include recurrence plot (RP), Gramian angular fields (GAF), Markov transition fields (MTF), and the methods proposed by Liu, Li, and Piugie. GAF, MTF, and RP were implemented using the Python package pyts, while the recent methodologies in Liu\cite{b25}, Li\cite{b27}, and Piugie\cite{b28} were implemented based on their respective literature. The implementation results are illustrated in Fig \ref{fig:allimages}. The configuration of deep SVDD remains consistent with the aforementioned description above.

\begin{table}[htbp]
\caption{ Performance comparison with other time-series to image encoding methods.}
\begin{center}
\begin{tabular}{|l|l|l|l|l|l|}
\hline
Model & EER & FAR & FRR & TAR & ACC \\
\hline
RP\cite{b22}     & 0.400 & \textbf{0.360} & 0.440 & 0.559 & 0.599 \\
GAF\cite{b23}   & 0.400 & 0.416 & 0.384 & 0.615 & 0.599 \\
MTF\cite{b23}    & 0.477 & 0.485 & 0.468 & 0.531 & 0.522 \\
Liu\cite{b25}    & 0.401 & 0.450 & 0.351 & 0.648 & 0.598 \\
Li \cite{b27}    & 0.424 & 0.729 & \textbf{0.118} & \textbf{0.881} & 0.575 \\
Piugie\cite{b28} & \textbf{0.353} & 0.366 & 0.347 & 0.652 & \textbf{0.646} \\
\hline
OURS(x,y) & 0.097 & 0.113 & 0.080 & 0.919 & 0.902 \\
OURS(PCA) & \textbf{0.067} & \textbf{0.080} & \textbf{0.055} & \textbf{0.944} & \textbf{0.932} \\
\hline
\end{tabular}
\label{tab:comparison}
\end{center}
\end{table}

Table \ref{tab:comparison} shows the performance of the time-series to image encoding methods on our dataset. Unlike other works, Piugie's image-based approach performed the best with an EER of 35.3\%, achieving an accuracy of 64.6\%. Despite each of them demonstrating state-of-the-art performance in their respective literature, employing a relatively simple deep learning model for user authentication led to a decrease in performance. This suggests that our approach captures temporal information better from keystrokes and emphasizes the importance of touch location information on a mobile device.

\section{Discussion}

In this section, we delve into feature engineering, the utilization of touch location data, and discussions regarding the collected dataset. Specifically, we propose directions for data pre-processing methods, highlight the limitations of existing timing-based data collection, and discuss methods for organizing a better dataset.

\subsection{Feature engineering}

Due to users' familiarity with mobile devices, they develop patterns when entering passwords such as PIN. The collected dataset revealed that the users' data followed a normal distribution. The conventional method of adjusting data to a specific range through min-max normalization may pose difficulties in recognizing user patterns. Conducting experiments on the user who exhibited the best performance in our system, as shown in Table \ref{tab:preprocessingcomparison}, the EER increased from 0.109 to 0.016 with standardization. This indicates that min-max normalization, which is sensitive to outliers, is not suitable to use for keystroke dynamics data due to the diverse distribution of user inputs. Therefore, pre-processing through standardization is preferable as the data follows a normal distribution.

\begin{table}[htbp]
\caption{ Experimental results based on pre-processing for the user whose result showed the best performance.}
\begin{center}
\begin{tabular}{|l|l|l|l|l|l|}
\hline
Pre-processing & EER & FAR & FRR & TAR & ACC \\
\hline
Min-max & 0.109 & 0.176 & 0.043 & 0.956 & 0.890 \\
Standardization & \textbf{0.016} & \textbf{0.010} & \textbf{0.025} & \textbf{0.975} & \textbf{0.982} \\
\hline
\end{tabular}
\label{tab:preprocessingcomparison}
\end{center}
\end{table}

\subsection{Touch location}

A model trained solely on timing values can be vulnerable to attacks such as shoulder surfing, especially when the mobile device is not on vibration mode, allowing the attacker to infer the timing by using the sound when the key is typed. On the other hand, in the case of touch location, estimating touch coordinates through visual observation is challenging unless a hidden application monitoring the screen is installed on the device. The Smudge attack, which infers touch coordinates using smudges left on the screen, is related to the user's finger movement patterns. However, obtaining precise touch coordinates through this method is practically challenging.

To assess the influence of user patterns in learning from touch location data and timing data, experiments were conducted by utilizing an autoencoder to remove each data component from standardized datasets. The results in Table \ref{tab:featureimpact} demonstrate better performance when only touch location data is utilized. Specifically, removing timing values resulted in an increase of 0.6\% in EER, while removing touch locations led to an increase of 3.9\% in EER. This emphasizes the significance of touch coordinates when implementing keystroke dynamics-based authentication systems on mobile devices.

\begin{table}[htbp]
\caption{ Experimental results based on feature set for the user whose result shows the best performance.}
\begin{center}
\begin{tabular}{|l|l|l|l|l|l|}
\hline
Dataset & EER & FAR & FRR & TAR & ACC \\
\hline
Original & 0.066 & 0.080 & 0.051 & 0.948 & 0.933 \\
\hline
\(-\) Location & 0.105 & 0.102 & 0.109 & 0.890 & 0.894 \\
\hline
\(-\) Timing & 0.072 & 0.070 & 0.075 & 0.924 & 0.927 \\
\hline
\end{tabular}
\label{tab:featureimpact}
\end{center}
\end{table}

\subsection{Dataset}

A comprehensive investigation was conducted on public datasets related to keystroke dynamics. However, due to limitations inherent in the data collection process for acquiring data from various devices, a public dataset containing touch location was notably absent. To address this gap, we independently collected data to create the desired feature dataset. The data collection took place in a laboratory environment, where participants sat without specific constraints, allowing them sufficient time to become familiar with the smartphone. However, touch habits may vary depending on a person's posture, behavior, or mood \cite{b30}. Although such information was not fully incorporated in the dataset we collected, further research is warranted to develop user profiles and implement user authentication systems capable of recognizing individual situations and applying our imaging methods accordingly.

\section{Conclusion}

In this paper, we propose a method for passive user authentication through keystroke dynamics encoded into images using deep SVDD. The purpose of image encoding is to differentiate imposter images from legitimate user images, aiding in effectively training the model to recognize user keystroke input patterns and enhancing imposter detection capabilities. Additionally, to achieve this, we collected keystroke dynamics datasets from 17 subjects along with touch location data, and standardization was employed instead of min-max normalization for data pre-processing based on the analysis of the dataset. Our system, leveraging touch information, achieved an average accuracy of 93.3\% and an EER of 0.06. The proposed encoding method shows enhanced user authentication performance compared to existing research relying on specialized models to achieve state-of-the-art performance. Furthermore, our approach is remarkably simple and intuitive, highlighting its substantial potential. As our method quietly performs passive authentication in the background from the fundamental PIN entry behavior, it offers scalability without imposing special actions on users while enhancing security.

Our future work includes expanding our system to cover not just keystroke dynamics like PINs, but also specialized behavior authentication like pattern authentication based on user touches. Our method shows effective for touch-based authentication systems with unique hidden patterns. Moreover, exploring data from diverse postures with tailored models could enhance performance by authenticating users using learned profile models for each posture.

\bibliographystyle{unsrt}  
\bibliography{references}

\begin{thebibliography}{10}

\bibitem{b1}
Ericsson.
\newblock Number of smartphone mobile network subscriptions worldwide from 2016 to 2022, with forecasts from 2023 to 2028.
\newblock \textit{Statista}, June 2023.
\newblock [Online] Available: \url{https://www.statista.com/statistics/330695/number-of-smartphone-users-worldwide/} [Accessed Mar 13, 2024].

\bibitem{b2}
Adam~J Aviv, Katherine Gibson, Evan Mossop, Matt Blaze, and Jonathan~M Smith.
\newblock Smudge attacks on smartphone touch screens.
\newblock In {\em 4th USENIX workshop on offensive technologies (WOOT 10)}, 2010.

\bibitem{b30}
Joseph Bonneau, Cormac Herley, Paul~C Van~Oorschot, and Frank Stajano.
\newblock The quest to replace passwords: A framework for comparative evaluation of web authentication schemes.
\newblock In {\em 2012 IEEE symposium on security and privacy}, pages 553--567. IEEE, 2012.

\bibitem{b3}
Emanuele Maiorana, Himanka Kalita, and Patrizio Campisi.
\newblock Mobile keystroke dynamics for biometric recognition: An overview.
\newblock {\em IET biometrics}, 10(1):1--23, 2021.

\bibitem{b7}
Neil~Zhenqiang Gong, Mathias Payer, Reza Moazzezi, and Mario Frank.
\newblock Forgery-resistant touch-based authentication on mobile devices.
\newblock In {\em Proceedings of the 11th ACM on Asia Conference on Computer and Communications Security}, pages 499--510, 2016.

\bibitem{b4}
Ioannis Stylios, Andreas Skalkos, Spyros Kokolakis, and Maria Karyda.
\newblock Bioprivacy: Development of a keystroke dynamics continuous authentication system.
\newblock In {\em European Symposium on Research in Computer Security}, pages 158--170. Springer, 2021.

\bibitem{b5}
Xiaofeng Lu, Shengfei Zhang, Pan Hui, and Pietro Lio.
\newblock Continuous authentication by free-text keystroke based on cnn and rnn.
\newblock {\em Computers \& Security}, 96:101861, 2020.

\bibitem{b6}
Anum~Tanveer Kiyani, Aboubaker Lasebae, Kamran Ali, Masood~Ur Rehman, and Bushra Haq.
\newblock Continuous user authentication featuring keystroke dynamics based on robust recurrent confidence model and ensemble learning approach.
\newblock {\em IEEE Access}, 8:156177--156189, 2020.

\bibitem{b8}
Benjamin Draffin, Jiang Zhu, and Joy Zhang.
\newblock Keysens: Passive user authentication through micro-behavior modeling of soft keyboard interaction.
\newblock In {\em Mobile Computing, Applications, and Services: 5th International Conference, MobiCASE 2013, Paris, France, November 7-8, 2013, Revised Selected Papers 5}, pages 184--201. Springer, 2014.

\bibitem{b10}
Maro Choi, Shincheol Lee, Minjae Jo, and Ji~Sun Shin.
\newblock Keystroke dynamics-based authentication using unique keypad.
\newblock {\em Sensors}, 21(6):2242, 2021.

\bibitem{b11}
Giuseppe Stragapede, Ruben Vera-Rodriguez, Ruben Tolosana, Aythami Morales, Alejandro Acien, and Ga{\"e}l Le~Lan.
\newblock Mobile behavioral biometrics for passive authentication.
\newblock {\em Pattern Recognition Letters}, 157:35--41, 2022.

\bibitem{b12}
Hanting Chen, Yunhe Wang, Tianyu Guo, Chang Xu, Yiping Deng, Zhenhua Liu, Siwei Ma, Chunjing Xu, Chao Xu, and Wen Gao.
\newblock Pre-trained image processing transformer.
\newblock In {\em Proceedings of the IEEE/CVF conference on computer vision and pattern recognition}, pages 12299--12310, 2021.

\bibitem{b13}
Kelei He, Chen Gan, Zhuoyuan Li, Islem Rekik, Zihao Yin, Wen Ji, Yang Gao, Qian Wang, Junfeng Zhang, and Dinggang Shen.
\newblock Transformers in medical image analysis.
\newblock {\em Intelligent Medicine}, 3(1):59--78, 2023.

\bibitem{b14}
Zekun Li, Shiyang Li, and Xifeng Yan.
\newblock Time series as images: Vision transformer for irregularly sampled time series.
\newblock {\em Advances in Neural Information Processing Systems}, 36, 2024.

\bibitem{b24}
Orcan Alpar.
\newblock Keystroke recognition in user authentication using ann based rgb histogram technique.
\newblock {\em Engineering Applications of Artificial Intelligence}, 32:213--217, 2014.

\bibitem{b25}
Mengxin Liu and Jianfeng Guan.
\newblock User keystroke authentication based on convolutional neural network.
\newblock In {\em Mobile Internet Security: Second International Symposium, MobiSec 2017, Jeju Island, Republic of Korea, October 19--22, 2017, Revised Selected Papers 2}, pages 157--168. Springer, 2019.

\bibitem{b26}
Anurag Tewari and Prabhat Verma.
\newblock An improved user identification based on keystroke-dynamics and transfer learning.
\newblock {\em Webology}, 19(1):5369--5387, 2022.

\bibitem{b27}
Jianwei Li, Han-Chih Chang, and Mark Stamp.
\newblock Free-text keystroke dynamics for user authentication.
\newblock In {\em Artificial Intelligence for Cybersecurity}, pages 357--380. Springer, 2022.

\bibitem{b28}
Yris Brice~Wandji Piugie, Jo{\"e}l Di~Manno, Christophe Rosenberger, and Christophe Charrier.
\newblock Keystroke dynamics based user authentication using deep learning neural networks.
\newblock In {\em 2022 International Conference on Cyberworlds (CW)}, pages 220--227. IEEE, 2022.

\bibitem{b15}
Daniel Buschek, Alexander De~Luca, and Florian Alt.
\newblock Improving accuracy, applicability and usability of keystroke biometrics on mobile touchscreen devices.
\newblock In {\em proceedings of the 33rd annual ACM conference on human factors in computing systems}, pages 1393--1402, 2015.

\bibitem{b16}
Kevin~S Killourhy and Roy~A Maxion.
\newblock Comparing anomaly-detection algorithms for keystroke dynamics.
\newblock In {\em 2009 IEEE/IFIP International Conference on Dependable Systems \& Networks}, pages 125--134. IEEE, 2009.

\bibitem{b17}
Esra Vural, Jiaju Huang, Daqing Hou, and Stephanie Schuckers.
\newblock Shared research dataset to support development of keystroke authentication.
\newblock In {\em IEEE International joint conference on biometrics}, pages 1--8. IEEE, 2014.

\bibitem{b18}
Yan Sun, Hayreddin Ceker, and Shambhu Upadhyaya.
\newblock Shared keystroke dataset for continuous authentication.
\newblock In {\em 2016 IEEE International Workshop on Information Forensics and Security (WIFS)}, pages 1--6. IEEE, 2016.

\bibitem{b19}
Zde{\v{n}}ka Sitov{\'a}, Jaroslav {\v{S}}ed{\v{e}}nka, Qing Yang, Ge~Peng, Gang Zhou, Paolo Gasti, and Kiran~S Balagani.
\newblock Hmog: New behavioral biometric features for continuous authentication of smartphone users.
\newblock {\em IEEE Transactions on Information Forensics and Security}, 11(5):877--892, 2015.

\bibitem{b20}
Sowndarya Krishnamoorthy, Luis Rueda, Sherif Saad, and Haytham Elmiligi.
\newblock Identification of user behavioral biometrics for authentication using keystroke dynamics and machine learning.
\newblock In {\em Proceedings of the 2018 2nd international conference on biometric engineering and applications}, pages 50--57, 2018.

\bibitem{b21}
Gabriel~Rodriguez Garcia, Gabriel Michau, M{\'e}lanie Ducoffe, Jayant~Sen Gupta, and Olga Fink.
\newblock Temporal signals to images: Monitoring the condition of industrial assets with deep learning image processing algorithms.
\newblock {\em Proceedings of the Institution of Mechanical Engineers, Part O: Journal of Risk and Reliability}, 236(4):617--627, 2022.

\bibitem{b22}
Jean-Pierre Eckmann, S~Oliffson Kamphorst, David Ruelle, et~al.
\newblock Recurrence plots of dynamical systems.
\newblock {\em World Scientific Series on Nonlinear Science Series A}, 16:441--446, 1995.

\bibitem{b23}
Zhiguang Wang and Tim Oates.
\newblock Encoding time series as images for visual inspection and classification using tiled convolutional neural networks.
\newblock In {\em Workshops at the twenty-ninth AAAI conference on artificial intelligence}, 2015.

\bibitem{b29}
Lukas Ruff, Robert Vandermeulen, Nico Goernitz, Lucas Deecke, Shoaib~Ahmed Siddiqui, Alexander Binder, Emmanuel M{\"u}ller, and Marius Kloft.
\newblock Deep one-class classification.
\newblock In {\em International conference on machine learning}, pages 4393--4402. PMLR, 2018.

\bibitem{b31}
John~V Monaco and Charles~C Tappert.
\newblock Obfuscating keystroke time intervals to avoid identification and impersonation.
\newblock {\em arXiv preprint arXiv:1609.07612}, 2016.

\end{thebibliography}

\end{document}